# Quality of Information in Mobile Crowdsensing: Survey and Research Challenges

FRANCESCO RESTUCCIA, Northeastern University, USA
NIRNAY GHOSH, SHAMEEK BHATTACHARJEE, and SAJAL K. DAS, Missouri S&T, USA
TOMMASO MELODIA, Northeastern University, USA

Smartphones have become the most pervasive devices in people's lives, and are clearly transforming the way we live and perceive technology. Today's smartphones benefit from almost ubiquitous Internet connectivity and come equipped with a plethora of inexpensive yet powerful embedded sensors, such as accelerometer, gyroscope, microphone, and camera. This unique combination has enabled revolutionary applications based on the mobile crowdsensing paradigm, such as real-time road traffic monitoring, air and noise pollution, crime control, and wildlife monitoring, just to name a few. Differently from prior sensing paradigms, humans are now the primary actors of the sensing process, since they become fundamental in retrieving reliable and up-to-date information about the event being monitored. As humans may behave unreliably or maliciously, assessing and guaranteeing Quality of Information (QoI) becomes more important than ever. In this paper, we provide a new framework for defining and enforcing the QoI in mobile crowdsensing, and analyze in depth the current state-of-the-art on the topic. We also outline novel research challenges, along with possible directions of future work.

## 1 INTRODUCTION

Smartphones have now become one of the most powerful pervasive technologies. According to the Ericsson Mobility Report of 2016, the number of smartphone subscriptions reached the staggering figure of 3.2 billion in 2015, and by the end of 2021, it is projected that there will be nearly 6.3 billion global mobile users, almost one for every person on the planet [45].

This work is partially supported by the NSF, under grants CNS-1545037, CNS-1545050, DGE-1433659 and IIS-1404673.
Authors' addresses: F. Restuccia and T. Melodia, Electrical and Computer Engineering Department, Northeastern University, 360 Huntington Ave, Boston, MA 02115, US. Email: {frestuc, melodia}@ece.neu.edu; N. Ghosh, S. Bhattacharjee and S.K. Das, Department of Computer Science, Missouri University of Science and Technology, 500 W 15th St, Rolla, MO 65409 US. Email: {ghoshn, shameek, sdas}@mst.edu.

Not only are today's smartphones ubiquitous devices, they are also equipped with a plethora of embedded multi-modal sensors; integrate wireless communication technologies such as 4G/WiFi Internet connectivity, and possess complex processing capabilities. For example, the cameras on smartphones can be used as video and image sensors [29], the microphone can be used as an acoustic sensor [39, 154, 206], and the embedded global positioning system (GPS) receiver can be used to gather accurate location information, while gyroscopes, accelerometers, and proximity sensors can be used to extract contextual information about the users, such as driving or walking states [75, 122]. Further, additional sensors can be easily interfaced with smartphones via Bluetooth or wired connections, such as temperature, air quality, and humidity [69, 168].

These technological features, combined with the advanced sensing capability of humans, has spurred a significant amount of research from both academia and industry, which together have proposed over the last 10 years a myriad of applications based on the emerging *mobile crowdsensing* paradigm[1]. Mobile crowdsensing empowers ordinary citizens (or users[2]) with the capability to actively monitor various phenomena pertaining to themselves (e.g., health, social connections) or their community (e.g., environment). This rich information or inference about themselves or the community may also be sent back to the participating or other concerned users to improve their life experiences, thus influencing their choices. Real-life applications, which can take advantage of both low-level sensor data and high-level user activities, range from real-time traffic monitoring applications, to environmental pollution monitoring, crime monitoring, and social networking, just to name a few. For a survey on mobile crowdsensing applications, we refer the reader to [92].

Although mobile crowdsensing allows the creation of disruptive technologies and applications, it also comes with novel and unique research challenges, which were absent in less complex sensing paradigms such as wireless sensor networks. The most important challenges that we will analyze in this paper can be summarized by the following three questions:

> *How can we trust that the smartphone-human sensors will send **useful** information?*
> *How can we **enforce** the submission of useful information?*
> *How can we **estimate** the usefulness of the submitted information?*

For each question, we have highlighted a key word that epitomizes a research challenge. Below, we briefly discuss each keyword and introduce the related research challenges that will be discussed in this paper.

- **Useful.** Mobile crowdsensing systems can be used for many purposes, all of them very different from each other and each of them requiring different constraints on the nature of information. For example, in a traffic monitoring application, information about traffic status on a given road should be refreshed every few minutes to allow motorists to arrange their routes in real time according to the traffic conditions. On the other hand, in an air pollution application, such sensing granularity may not be necessary, as information regarding pollution usually may not change as often. In such diverse and complex context, is it possible to define the notion of *useful information*, also referred to as *quality of information* (QoI)? Specifically, what are the characteristics that define QoI for the wide variety of the mobile crowdsensing systems nowadays available? Can traditional definitions of QoI, for example, in sensor networks [17], be applied to mobile crowdsensing?

---

[1]Henceforth, we will refer to the term *mobile crowdsensing* to designate applications where participants voluntarily contribute sensor data for their own benefit or for the benefit of the community by using their phones. Such a notion therefore includes participatory sensing [20], mobiscopes [1], opportunistic sensing [23], and equivalent terms such as mobile phone sensing and smartphone sensing [98], or mobile crowdsensing [65]. It also covers specific terminologies such as urban sensing [23], citizen sensing [20], people-centric sensing [23], [24], and community sensing [97].

[2]In the following, we will use the terms "participants" and "users" interchangeably, as well as "system", "application" and "platform".

- **Enforce.** Once the notion of QoI has been defined, a mobile crowdsensing system has to provide a set of minimum QoI requirements that need to be satisfied. For example, an air pollution application may require to receive 100 sensing reports from 10 different sources in a given neighborhood of a city during an hour. In traditional sensor networks, this challenge could be "easily" overcome by deploying more sensors and making sure constraints on the bandwidth and packet delivery rate are satisfied, usually by solving complex optimization problems [43]. However, in mobile crowdsensing the control of the sensing process is mainly delegated to *people*, who may choose to deliver (or not) relevant information depending on their *willingness* and personal *constraints*, such as time, smartphone battery, available network bandwidth, etc. Other factors may also influence the QoI of submitted sensing reports. For example, smartphone sensors are not typically of the same fidelity as task-specific sensors (e.g., a noise pollution monitor vs. smartphone microphone). Also, inadvertent erroneous handling of the phone (e.g., capturing blurry photos by mistake), or lack of experience in performing the sensing tasks may adversely influence the QoI, irrespective of the willingness of the participants. Moreover, the degree of participation is also heavily influenced by the participants' current *mobility pattern*, which frequently changes over time as most users of mobile crowdsensing systems are pedestrian, drivers, or people commuting to their workplace [141]. Due to the strong human component in the sensing process, enforcing QoI in mobile crowdsensing is a peculiar issue and requires considerable research effort.

- **Estimate.** After having received the sensing information from the participants, but before using it, the crowdsensing system must be able to make sure the information contained in the sensing reports is actually compliant to the QoI constraints required by the sensing application. However, it is significantly challenging to estimate the QoI of sensing reports submitted by smartphone-human sensors *in the wild*. First, smartphone sensors could be faulty, uncalibrated, or tampered with. Second, some participants may exhibit *malicious* behavior, trying to disrupt the sensing application by deliberately sending low-QoI information. For example, in 2014 a group of students from Technion-Israel Institute of Technology successfully simulated through GPS spoofing a traffic jam on the *Waze* [194] road traffic monitoring application, currently used by more than 50 million people worldwide [173]. This simulation lasted hours, causing thousands of motorists to deviate from their planned routes [7]. Third, the information contained in sensing reports coming from the same geographical area at (almost) the same time could be highly *conflicting*, further complicating the QoI estimation process as traditionally used techniques such as sensor fusion might become ineffective [101]. The unique combination of these aspects makes the issue of estimating QoI in mobile crowdsensing a significantly challenging process.

To the best of our knowledge, existing literature lacks a framework encompassing and surveying the main aspects of QoI for mobile crowdsensing systems, including the definition of existing and future research challenges in this field. For this reason, within the scope of this manuscript, we analyze the state-of-the-art in QoI as applied to mobile crowdsensing campaigns; besides describing the solutions currently applied, we also highlight their limitations, and discuss open issues as well as their impact on QoI sensing surveys. Specifically, our novel contributions can be summarized as follows. First, by surveying the most relevant mobile crowdsensing applications currently available, we propose and discuss a unified framework that defines the concept of "quality of information" (QoI) in mobile crowdsensing systems. Second, we define and analyze in detail the research challenges in enforcing and estimating the QoI in mobile crowdsensing systems, as well as discussing the already existing related work on the topic. Finally, by analyzing the limits and weaknesses of existing work, we provide a roadmap of possible directions of novel future research that we hope will inspire additional research in this exciting field.

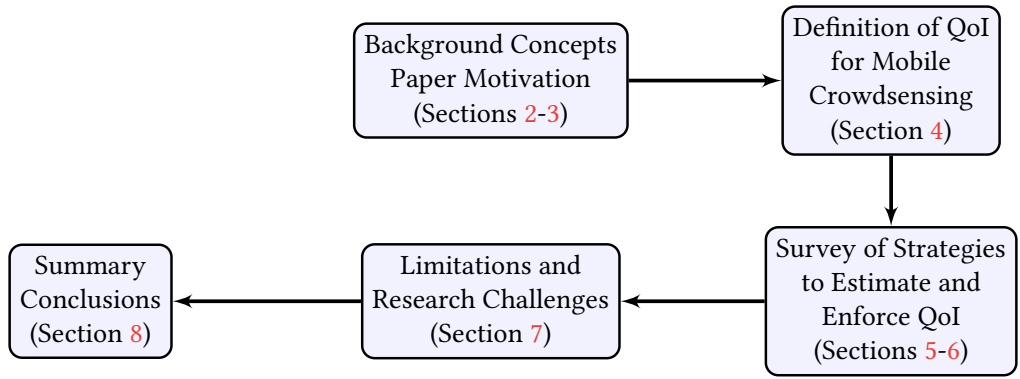

Fig. 1. Paper organization - A top-down approach to QoI.

Figure 1 outlines the organization of the paper – we will present our work by following a top-down approach. Section 2 provides background information on mobile crowdsensing and briefly surveys related work on QoI. Section 3 summarizes the motivations of this work by highlighting the differences between mobile crowdsensing and prior sensing paradigms. Section 4 proposes a novel framework that outlines the operations needed to define, enforce and estimate QoI in mobile crowdsensing systems. We then survey the state of the art pertaining to each component of our framework in Sections 5 and 6. At the end of these two sections, we discuss in details the limitations of existing work and provide suggestions on how to improve it. Section 7 presents open research challenges toward improving the QoI in mobile crowdsensing, while Section 8 concludes the paper.

## 2 WHAT IS MOBILE CROWDSENSING?

In this section, we briefly introduce the concept of mobile crowdsensing and discuss related work on QoI in other fields different from mobile crowdsensing.

Mobile crowdsensing is a paradigm that was first proposed in [20]. In a nutshell, it leverages the widespread usage of smartphones to acquire up-to-date and fine-grained information about a location (or event) of interest. As far as the architectural components are concerned, most of the existing mobile crowdsensing systems are generally composed of the *participants*, the *sensing application* (app), and the *mobile crowdsensing platform* (in short, MCP).

- The core component of the whole architecture are the *participants*, whose task is to use their smart devices to capture different kinds of sensor data, such as location, images, sound samples, accelerometer data, biometric data, and barometric pressure, just to name a few. However, more complex information about the sensing area or an environmental phenomenon, such as traffic status or weather information, may also be provided. In practical implementations, participants typically register with the system by providing a username and password [194] that allow their contributions to be uniquely identified.
- The *sensing application* (app), deployed on the users' smartphones, is distributed through common application markets like *Google Play* or *App Store*, or is retrieved from a mobile cloud computing system [46]. It is responsible for providing the users with a friendly interface for data acquisition and visualization. In particular, data acquisition may be triggered by the users themselves or elicited by the app, on a one-time on-demand basis or periodically.
- The backend component of the system, and the place where most of the computation is executed, is the *mobile crowdsensing platform* (in short, MCP). This platform is usually responsible for filtering, elaboration, and redistribution of sensed data, as well as coordinating every operation performed by the system. This component is usually implemented by a set of servers dedicated to the processing of sensed data [144]. Furthermore, the MCP also ensures efficient storage and elaboration of the sensed data coming from the users, which may be stored in some relational databases [60], or databases specially adapted to the management of sensor readings. Along with the main elaboration system, the MCP might leverage a *reputation system* and an *incentive mechanism*. Briefly, the target of a reputation system is to predict the reliability of the data sent by the users based on their past behavior [78, 188] so as to filter out unreliable reports. Conversely, the target of an incentive mechanism is to encourage user participation by appropriately rewarding the users for their contributions to the mobile crowdsensing campaign [147].

## 3 WHY IS THE QOI PROBLEM IN MOBILE CROWDSENSING SO CHALLENGING?

Traditionally, the notion of quality in sensor networks has been typically thought of as a derivative of the traditional quality of service (in short, QoS) in networks, typically assessed along various QoS attributes, such as bandwidth, delay, delay jitter, and packet loss probability. Therefore, the "quality" level of a network translates into a judgement statement about the network's ability to attain various levels for these attributes that are appropriate (hence of value) for a particular class of applications, such as best effort and constant or variable bit-rate.

Analogously, one would expect that the "goodness" of a mobile crowdsensing system in performing its tasks of producing and providing sensory information to stakeholders would relate to QoS attributes, and how well the values of these attributes reflect and accommodate the information needs of applications that consume it. However, these *network-oriented* quality aspects of sensor networks do not capture in their entirety the *application-oriented* quality aspects of the system.

To make this point, we report a citation by Peter Drucker, first published in his book "Innovation and Entrepreneurship" (1985). Focused on customer satisfaction, he defined quality as follows:

> "Quality in a product or service is not what the supplier puts in.
> It is what the customer gets out and is willing to pay for."

From this simple yet very profound statement, we can derive that quality is something which is usually *perceived* by the final user, and not what the object is *per se*. By extending this definition to the context of mobile crowdsensing, we derive that the information generated in mobile crowdsensing systems becomes of value (and therefore, has quality) only when the system (or a portion of it) *depends on* that information, and in particular, *needs* that information to implement its functionality. Therefore, the standard definition of QoS for networks cannot be applied to define the QoI for mobile crowdsensing systems.

Furthermore, there are fundamental differences between mobile crowdsensing and other paradigms, such as traditional crowdsourcing and wireless sensor networks. Below we summarize why QoI in mobile crowdsensing cannot be defined and attained by traditional methodologies used in other fields of computer networking.

- *Mobile crowdsensing systems handle extremely diverse types of information.* Since its inception in 2006, a plethora of different mobile crowdsensing systems have been proposed for many different purposes, from emotional and health monitoring [134, 139] to noise mapping systems [39, 136], road traffic monitoring [155, 210], discovery of people in distress [2, 148], and so on. Each system uses different kinds of sensors and may or may not require interaction with the participants. Therefore, traditional definitions of QoI hardly apply to mobile crowdsensing systems. Therefore, the QoI definition must include a broader and more comprehensive view of QoI to entail the variety of applications currently available.

- *We can't assume there will be a steady flow of information received by the system.* In traditional sensing paradigms, it is reasonable to assume the system will obtain a steady amount of information generated by the network in a given time frame. This may be achieved by leveraging QoS-aware algorithms and protocols for medium channel access [64, 145, 146], node discovery [96, 142, 143], routing to the sink node [171], and topology management [203], among other strategies. Given in mobile crowdsensing humans become the primary actors of the sensing process, it is not possible to directly apply QoS-aware algorithms to obtain a steady flow of sensed data. This is because such protocols do not account for the fact that humans may choose to deliver (or not) the information depending on their *willingness* and *constraints*, such as smartphone battery, available network bandwidth, and time to perform the sensing service. Humans are also *highly mobile*, meaning that their trajectory may not be known a priori and it is rarely controllable [12]. This aspect is fundamentally different from traditional crowdsourcing, where human participants are not generally assumed to be mobile (e.g., Amazon Mechanical Turk). This implies that novel mechanisms must be designed to guarantee a steady flow of information to the system.

- *We can't assume the information collected by the system will be accurate.* An abundance of research work has been dedicated to address the problem of data accuracy, usually through data aggregation [72, 151, 163] and trust management [9, 68, 100]. Although these solutions are significantly effective in the context of sensor networks, it is intuitive that the same techniques may not be directly applied to the context of mobile crowdsensing. There are several reasons that demonstrate this point. First, task-specific sensors deployed in sensor networks have far better *fidelity* than the sensors deployed on smartphones, which are designed for simple tasks such as phone orientation or basic activity recognition (e.g., walking). Second, humans may *lack expertise* or willingness when performing the sensing tasks. For example, participants may inadvertently capture blurry photos or place their smartphones in a noisy environment, thus invalidating their effort in providing meaningful sensor data.

  Another crucial issue in mobile crowdsensing, also studied in the context of generic crowdsourcing, is how to deal with *malicious behavior* by participants [33, 54, 179]. Indeed, conversely from prior paradigms, in mobile crowdsensing it becomes extremely simple for malicious users to disrupt the sensing application by sending low-QoI information. Most often, malicious participants are motivated by the desire of earning a reward with little or no effort [53], spamming [150], or influencing other participants' behavior [7, 71]. The practical consequence of malicious and unreliable behavior is that the information received by the system will be highly *conflicting*, further complicating the QoI estimation process as traditionally used techniques such as sensor fusion might become ineffective [101].

## 4 A COMPREHENSIVE FRAMEWORK FOR QOI IN MOBILE CROWDSENSING

Given the wide variety of mobile crowdsensing applications, we need a notion of quality of information (QoI) flexible and broad enough to entail different applications and existing works on QoI proposed so far in the literature. This section formulates a coherent and comprehensive QoI framework, as well as briefly discusses related work on QoI.

**Definition.** The Quality of Information (QoI) in mobile crowdsensing is:

(1) the set of constraints specified by the system in terms of Information Quantity and Information Accuracy, collectively called *QoI Constraints*;
(2) the set of activities in the *QoI Loop* needed to achieve the QoI Constraints defined by the system, hereafter referred to as *QoI Enforcement* and *QoI Estimation*.

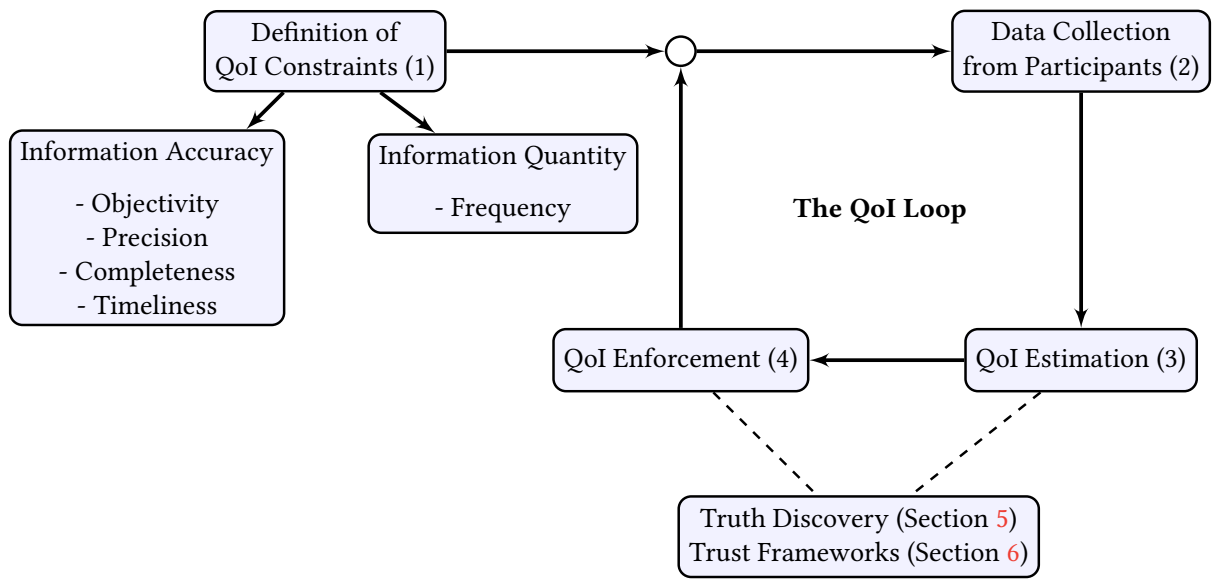

Fig. 2. The "QoI loop" of mobile crowdsensing systems.

With the help of the flowchart in Figure 2, let us now discuss in details each of the concepts pertaining QoI.

(1) Before deployment, the system has to clearly specify the QoI constraints needed by the application. Such constraints must be clearly quantified in objective (possibly, numeric) terms. These constraints depend on the type of data the system is collecting, and the application needs. In details, we define Information Accuracy and Quantity as follows:
    - As in [8], we use the term *Information Accuracy* to indicate "the correctness of the information" received by the system. Specifically, we will use this term to encompass both the elements of intrinsic QoI, which are **(i)** objectivity and **(ii)** precision, and contextual QoI, which are **(iii)** completeness and **(iv)** timeliness [185]. Although there might be other attributes to describe information accuracy, we believe these four attributes are sufficient for the scope of existing mobile crowdsensing applications.
    - *Information Quantity* defines the amount of information that is received by the system over a period of time, in other words, the *frequency* of information retrieval [17]. The constraint on Information Quantity must be specified in numerical terms, e.g., in terms of sensing reports received in an hour. Such constraint may also vary over time (e.g., more data is needed during day time than during the night) and over space (e.g., some neighborhoods are more important than other).

(2) The first (and most fundamental step) in the QoI loop is collecting sensing reports from the participants. Given that mobile crowdsensing relies completely on the voluntary participation of citizens, recruiting and incentivizing users' active participation becomes a key factor. For this reason, a number of incentive mechanisms have been proposed to increase not only the collected amount of data, but also its quality. Furthermore, a number of frameworks aimed to recruit participants in order to maximize the coverage of the sensing area have been recently proposed [30, 61, 85, 86, 104, 182, 183, 208], including detailed reviews of the related literature [58, 81, 147]. Therefore, in this paper we focus on the QoI Estimation and Enforcement steps of the QoI loop, which has received relatively less attention.

(3) After the data collection phase, but before accepting the information into the system, the QoI of the collected data must be estimated and meaningful information must be extracted from the crowdsourced data. This step (called *QoI Estimation*) is crucial to make sure that the data collected by the system actually respects the QoI constraints defined by the application. After assessing its QoI level, the information can be redistributed to the concerned participants, by using websites or the sensing app used to transmit the sensed information [57].

(4) After identifying the QoI constraints, the *QoI Enforcement* step must take place to make sure that the QoI constraints are respected and useful information is inferred from the (most likely conflicting) sensing reports reported by the participants. Usually, these tasks are accomplished with the help of truth discovery [101] or trust frameworks [118]. Given their paramount importance in the QoI loop, the rest of the paper will be devoted to the discussion of existing truth discovery (Section 5) and trust frameworks (Section 6).

*4.0.1 Related work.* Defining and modeling the concepts of quality and quality of information (QoI) has been (and still is) the objective of much work throughout the years. This effort is justified by the fact that information has always been considered to be an asset that a business unit can plan and exercise control over. This asset aspect of information has given rise to management processes for ensuring QoI [44], as well as to the industry efforts to establish good data quality management practices for inter- and intra-business processes [13]. Another example is the ISO 9001 quality standard, which was first proposed in 1987 but is still used to determine quality in management systems. Specifically, ISO 9000 [125] is a family of standards related to the quality management systems. It defines quality (within its context) as the "degree to which a set of inherent characteristics fulfills requirements".

As far as the computerized world is concerned, QoI has been traditionally studied in the context of enterprise data management systems and processes, the outcome of Web searches [95] and with information fusion applications of detection, classification, identification, and tracking for military applications [34]. The most used definition of QoI in these cases refers to information that is "fit-to-use" for a purpose [42]. A plethora of research has also been conducted in the telecommunication and networking domains to define and enforce QoI. For example, the ITU-T Rec. E.800 [170] deals with QoS provisioning of digital telecommunication services. Similarly, the IETF RFC 2386 [49] deals with a framework for QoS-based routing in the Internet.

In the sensing domain, the first paper to define and deal with QoI in a holistic way is [17]. In this article, the authors highlight, refine, and extend a two-layer definition of QoI, where the first relates to context-independent aspects and the latter to context-dependent aspects of an information product. They also introduce a framework for scoring and ranking information products based on their value attributes. Building on this novel definition, much work has since been proposed to address QoI in networking [28, 51, 111, 164].

## 5 TRUTH DISCOVERY

In this section, we survey the state of the art in truth discovery in mobile crowdsensing, and also propose novel research challenges. First, in Section 5.1 we provide an overview of truth discovery, formalize its problem definition, and discuss related work. Next, in Section 5.2 we discuss background information necessary to grasp the main concepts of existing truth discovery frameworks, which are subsequently discussed in detail in Section 5.3. We then conclude the section by discussing how to improve existing work in Section 5.4.

### 5.1 Overview, Problem Definition and Related Work

The field of truth discovery has recently attracted enormous attention from the data mining and crowdsourcing communities [102, 175]. This stems from the fact that the information sent by two or more participants about the same event is likely to be conflicting. As discussed earlier, this may be due to several reasons, such as erroneous handling of the smartphone, noisy environment, typing error, or network delay, and so on.

To explain this point, let us assume that three participants, also called *sources* in the lexicon of truth discovery, are reporting information regarding the gas price at a gas station at a particular time instance. The reports submitted by the participants contain the following information: $2.89, $2.57, and $2.85. The fundamental question that one may ask is which quantity best represents the real status of the event (in this case, the gas price at that particular station), and also what led the participants to report an unreliable piece of information. In this and many more sensing scenarios, it is essential to *aggregate noisy information* about the same event collected from different participants. One approach could be to use simple strategies, such as majority voting or averaging of the sensing reports. However, such strategies may not be effective as they do not consider that participants may not be equally reliable, or they may collude.

For example, referring to the previous gas station example, we know that one report may be more reliable than the other as it comes from the owner of the gas station. Therefore, the accuracy of aggregated results can be improved by capturing the reliabilities of participants. The challenge is that the reliability of participants is usually unknown a priori and has to be somehow inferred.

Let us now introduce some definitions and notations that will be used throughout this section for clear and consistent discussions.

- *Entity (or object)*: An entity $j \in \mathcal{E}$ is the object or quantity being monitored. This can be either air quality measurement, or traffic situation at a particular location, or presence of potholes on a given street, and so on;
- *Source (or participant)*: A source $i \in \mathcal{S}$ generates a sensing report regarding an entity. Thus, if entity $j$ is observed by source $i$, we denote the report as $x_{i,j}$;
- *Source-claim matrix*: A source-claim matrix $SC$ is of the order $\mathcal{S} \times \mathcal{E}$. Each entry $S_iC_j$ keeps track of whether a source $i$ has made any claim for the entity $j$;
- *Reliability (or weight)*: The reliability $w_i$ of participant $i$ is a quantity that expresses the probability of participant $i$ to provide reliable information, where higher reliability indicates that $i$ is more reliable and the information is more likely to be accurate;
- *Truth*: The truth (also truth value) for an entity $j$, defined as $x_j^*$, is the information selected as the most reliable among the set of candidate observations submitted by different sources.

Based on these definitions, we define the general truth discovery problem as follows.

DEFINITION 1. **Truth Discovery Problem.** *Let $\mathcal{S}$ be a set of sources and $\mathcal{E}$ the set of entities being monitored. For each entity $j \in \mathcal{E}$, find an optimal truth value $\{x_j^*\}_{j \in \mathcal{E}}$, from a set of observations $\{x_{i,j}\}_{i \in \mathcal{S}, j \in \mathcal{E}}$ submitted by different sources.*

*5.1.1 Related work.* Research on truth discovery from the machine learning and data mining communities has provided several interesting results. The first efforts on the topic attempted to solve a similar problem in information networks using heuristics whose inspiration can be traced back to Google's *PageRank* [19]. PageRank iteratively ranks the credibility of sources on the Web, by iteratively considering the credibility of sources who link to them.

Extensions of PageRank, known as *fact-finders*, iteratively compute the credibility of sources and claims. Specifically, they estimate the credibility of claims from the credibility of sources that make them (source reputation estimation), then estimate the credibility of sources based on the credibility of their claims (truth computation step). More formally, the reliability (weight) $w_i$ of participant $i$ at the $k^{th}$ step of the iterative algorithm is calculated in terms of the trustworthiness of its claims on entity $j$ in the previous iteration $x_{i,j}^{k-1}$, and the truth of each claim $x_j^*$ in terms of $\{w_i\}_{i \in \mathcal{S}}$. The truth computation step and source reputation estimation are repeated until convergence. A general truth discovery algorithm is provided in Algorithm 1.

---

**ALGORITHM 1:** General Framework for Truth Discovery.

**Input**: Observations from sources $\{x_{i,j}\}_{i \in \mathcal{S}, j \in \mathcal{E}}$
**Output**: Optimal truths $\{x_j^*\}_{j \in \mathcal{E}}$ and the estimated source reliability $\{w_i\}_{i \in \mathcal{S}}$
**Initialize**: Source weights $\{w_i\}_{i \in \mathcal{S}}$;
**while** *Convergence not achieved* **do**
    **for** *each* $j \in \mathcal{E}$ **do**
        **Truth computation**: infer truth for entity $j$ based on the current estimation of source weights;
    **end**
    **Source weight estimation**: update source weights $\{w_i\}_{i \in \mathcal{S}}$ based on the current identified truths;
**end**
**Return**: $\{x_j^*\}_{j \in \mathcal{E}}, \{w_i\}_{i \in \mathcal{S}}$;

---

One of the first efforts in this sense was proposed in [202], where the authors introduced TruthFinder as an unsupervised fact-finder for trust analysis on a providers-facts network. TruthFinder calculates the "probability" of a claim by assuming that each source's reputation is the probability of it being correct, and then averages reputations to obtain trustworthiness scores for each claim. In details,

$$w_i^k = \frac{\sum_{j \in S_i C_j} x_{i,j}^{k-1}}{|\{S_i C_j\}_{j \in \mathcal{E}}|} \qquad x_j^k = 1 - \prod_{i \in S_i C_j}(1 - w_i^k) \qquad (1)$$

Similarly, Pasternack *et al.* extended such framework in 2010 by proposing three algorithms on fact-finding, namely *Average.Log, Investment, and Pooled Investment* [128]. Other fact-finders enhanced the basic framework by incorporating analysis on properties or dependencies within assertions or sources [55, 132, 174]. For example, Galland *et al.* took the notion of hardness of facts into consideration by proposing the algorithms *Cosine, 2-Estimates, and 3-Estimates* [55]. The Bayesian Interpretation scheme presented by [174] provided an approximation approach to truth estimation. This interpretation lead to a general foundation for using information network analysis to quantify, in an analytically-founded manner, the probability that each fact or source is correct. However, this approach is very sensitive to initial conditions of iterations, as pointed out by [178]. More recently, new fact-finding algorithms have been designed to address the background knowledge [129], multi-valued facts [209], and multi-dimensional aspects of the problem [204].

There exists a significant amount of literature in the machine learning community to improve data quality and identify low quality labelers in a multi-labeler environment. For a comprehensive survey, the reader may refer to [50]. Sheng *et al.* proposed in [157] a repeated labeling scheme to improve label quality by selectively acquiring multiple labels and empirically comparing several models that aggregate responses from multiple labelers. Dekel *et al.* applied a classification technique to simulate aggregate labels and prune low-quality labelers in a crowd to improve the label quality of the training dataset [36]. However, all of the above approaches made explicit or implicit assumptions that may not be applicable for mobile crowdsensing applications. For example, [157] assumed labelers were known a priori and could be explicitly asked to label certain data points. Moreover, [36] assumed most of labelers were reliable and the simple aggregation of their labels would be enough to approximate the ground-truth. In contrast, participants in social sensing usually upload their measurements based on their own observations and the simple aggregation technique (e.g., majority voting) was shown in [128] to be inaccurate when the reliability of participant is not sufficient. More recently, in [47] Fang *et al.* formulated the problem of improving labeling quality by characterizing the knowledge set of each labeler. Tarasov *et al.* proposed in [166] an algorithm called dynamic estimation of rater reliability in regression (DER3), which dynamically estimates the reliability of participants and accepts only those participants who are deemed to be reliable.

Finally, maximum likelihood estimation (MLE) [119, 130, 158, 184] and Kalman Filtering (KF) [110, 123, 190] frameworks have been widely used in the wireless sensor networks (WSNs) and data fusion communities. For example, Pereira *et al.* proposed in [130] a diffusion-based MLE algorithm for distributed estimation in WSN in the presence of data faults. Sheng *et al.* developed in [158] a MLE method to infer locations of multiple sources by using acoustic signal energy measurements. Msechu *et al.* proposed in [119] a MLE based approach to aggregate the signals from remote sensor nodes to a fusion center without any inter-sensor collaborations. Olfati-Saber proposed in [123] three algorithms based on KF that use consensus filters and covariance information for fusion of the sensor data. Masazade *et al.* study the problem of target tracking based on energy readings of sensors in [110], and propose a framework to minimize the estimation error of sensors by using a Kalman filter. The same problem is tackled by Wang *et al.* in [190], where the authors use the combined effect of MLE and KF to remove the sensing nonlinearity and obtain better tracking performance. The above work on MLE and KF primarily focused on the estimation of continuous variables from physical sensor measurements, and does not take into account any contextual information, e.g., the reputation of sources.

## 5.2 Background Concepts

We briefly provide some background on mathematical models and concepts that will be used throughout this section.

### 5.2.1 Maximum Likelihood Estimation (MLE).

Suppose there is a sample $X = \{x_1 \cdots x_n\}$ of independent and identically distributed (i.i.d.) observations, coming from a distribution with an unknown probability density function (p.d.f.) $f(\cdot)$. It is assumed that $f(\cdot)$ belongs to a family of distributions $f(\cdot|\theta), \theta \in \Theta$ called the *parametric model*, where $\theta$ is a vector of parameters for this family such that $f = f(\cdot \mid \theta)$. The value of $\theta$ is unknown and is referred to as the *true value* of the parameter vector. It is desirable to find an estimator $\hat{\theta}$ which would be as close to the true value $\theta$ as possible. Either or both the observed variables $x_i$ and $\theta$ can be vectors.

To use the method of maximum likelihood, the densities for all observations must be specified. For an i.i.d. sample, this joint density function is

$$f(x_1, x_2, \ldots, x_n \mid \theta) = f(x_1 \mid \theta) \times f(x_2|\theta) \times \cdots \times f(x_n \mid \theta). \tag{2}$$

Now let us look at this function from a different perspective by considering the observed values $X$ to be fixed "parameters" of this function, whereas $\theta$ will be the function's variable and allowed to vary freely; this same function will be called the likelihood function:

$$\mathcal{L}(\theta\,;x_1,\ldots,x_n) = f(x_1, x_2, \ldots, x_n \mid \theta) = \prod_{i=1}^{n} f(x_i \mid \theta). \tag{3}$$

Note that the semicolon denotes a separation between the two categories of input arguments: the parameters $\theta$ and the observations $X$. In practice the algebra is often more convenient when working with the natural logarithm of the likelihood function, called the "log-likelihood":

$$\ln \mathcal{L}(\theta\,;x_1,\ldots,x_n) = \sum_{i=1}^{n} \ln f(x_i \mid \theta), \tag{4}$$

or the average log-likelihood $\hat{\ell} = (1/n) \ln \mathcal{L}$. The hat over $\ell$ indicates that it is akin to some estimator. Indeed, $\hat{\ell}$ estimates the expected log-likelihood of a single observation in the model. The method of maximum likelihood estimates $\theta$ by finding a value that maximizes $\hat{\ell}(\theta; x)$. This method of estimation defines a *maximum likelihood estimator* (MLE) of $\theta$:

$$\{\hat{\theta}_{\mathrm{mle}}\} \subseteq \{\arg\max_{\theta \in \Theta} \hat{\ell}(\theta; x_1, \ldots, x_n)\}, \tag{5}$$

if a maximum exists. An MLE estimate is the same regardless of whether we maximize the likelihood or the log-likelihood function, since log is a monotonically increasing function. Above, it is assumed that the data are independent and identically distributed. The method can be applied however to a broader setting, as long as it is possible to write the joint density function $f(x_1, x_2, \ldots, x_n \mid \theta)$, and its parameter $\theta$ has a finite dimension which does not depend on the sample size $n$.

An MLE coincides with the most probable Bayesian estimator given a uniform prior distribution on the parameters. Indeed, the maximum a posteriori estimate is the parameter $\theta$ that maximizes the probability of $\theta$ given the data:

$$P(\theta \mid x_1, x_2, \ldots, x_n) = \frac{f(x_1, x_2, \ldots, x_n \mid \theta) \cdot P(\theta)}{P(x_1, x_2, \ldots, x_n)} \tag{6}$$

where $P(\theta)$ is the prior distribution for the parameter $\theta$ and where $P(x_1, x_2, \ldots, x_n)$ is the probability of the data averaged over all parameters. Since the denominator is independent of $\theta$, the Bayesian estimator is obtained by maximizing $f(x_1, x_2, \ldots, x_n \mid \theta) \cdot P(\theta)$ with respect to $\theta$.

For many models, an MLE can be found as an explicit function of the observed data $X$. For many other models, however, no closed-form solution to the maximization problem is known or available, and an MLE has to be found numerically using iterative mechanisms, such as expectation maximization (EM). For some problems, there may be multiple estimates that maximize the likelihood. For other problems, no maximum likelihood estimate exists and either the log-likelihood function increases without ever reaching a supremum value, or that the supremum does exist but is outside the bounds of $\Theta$, the set of acceptable parameter values.

To make an example with Bernoulli variables, let us suppose that $X = x_1 \cdots x_n$ represents the outcomes of $n$ independent Bernoulli trials, each with success probability $p$. Since the observations are taken from i.i.d. random variables, the distribution is

$$\mathcal{L}(p; x_1, \ldots, x_n) = \prod_{i=1}^{n} f(x_i; p) = \prod_{i=1}^{n} p^{x_i}(1-p)^{1-x_i} = p^{\sum_i x_i}(1-p)^{n-\sum_i x_i} \tag{7}$$

The log-likelihood is therefore:

$$\ln \mathcal{L}(p; x_1, \ldots, x_n) = \sum_{i=1}^{n} x_i \cdot \ln p + (n - \sum_{i=1}^{n} x_i) \cdot \ln(1-p) \tag{8}$$

Differentiating $\ln \mathcal{L}(p; x_1, \ldots, x_n)$ with respect to $p$ and setting the derivative to zero shows that this function achieves a maximum at $\hat{p} = \sum_{i=1}^{n} x_i / n$. Since $\sum_{i=1}^{n} x_i$ is the total number of successes observed in the $n$ trials, $\hat{p}$ is the observed proportion of successes in the $n$ trials.

*5.2.2 Expectation Maximization (EM).* Expectation maximization is a general optimization technique for finding the maximum likelihood estimation of parameters in a statistical model, where the model depends on unobserved latent variables [37]. It iterates between two main steps (namely, the E-step and the M-step) until the estimation converges (i.e., the likelihood function reaches the maximum).

(1) *E-step*: Calculate the expected value of the log-likelihood function, with respect to the conditional probability distribution of $\mathbf{Z}$ given $\mathbf{X}$ under the current estimate of the parameters $\theta^{(t)}$:
$$Q(\theta|\theta^{(t)}) = \mathrm{E}_{\mathbf{Z}|\mathbf{X},\theta^{(t)}}\left[\log L(\theta;\mathbf{X},\mathbf{Z})\right]$$

(2) *M-step*: Find the parameter $\theta^{(t+1)}$ that maximizes the quantity
$$\theta^{(t+1)} = \arg\max_{\theta} Q(\theta|\theta^{(t)})$$

*5.2.3 Maximum-A-Posteriori (MAP) Estimation.* Assume that we want to estimate an unobserved population parameter $\theta$ on the basis of observations $x$. Let $f$ be the sampling distribution of $x$, so that $f(x|\theta)$ is the probability of $x$ given the underlying population parameter is $\theta$. Then the likelihood function is given as:
$$\theta \longmapsto f(x|\theta) \qquad (9)$$
The maximum likelihood estimate of $\theta$ is as follows:
$$\hat{\theta}_{ML}(x) = \arg\max_{\theta} f(x|\theta) \qquad (10)$$
Let a prior distribution $g$ over $\theta$ exists. This allows us to treat $\theta$ as a random variable as in Bayesian statistics. We can calculate the posterior distribution of $\theta$ using Bayes' theorem:
$$\theta \longmapsto f(\theta|x) = \frac{f(x|\theta) \cdot g(\theta)}{\int_{\vartheta \in \Theta} f(x|\vartheta) \cdot g(\vartheta) d\vartheta} \qquad (11)$$
where $g$ is the density function of $\theta$, and $\Theta$ is the domain of $g$.

The method of maximum a posteriori estimation then estimates $\theta$ as the mode of the posterior distribution of this random variable:
$$\hat{\theta}_{ML}(x) = \arg\max_{\theta} f(\theta|x) = \frac{f(x|\theta) \cdot g(\theta)}{\int_{\vartheta} f(x|\vartheta) \cdot g(\vartheta) d\vartheta} = \arg\max_{\theta} f(x|\theta) \cdot g(\theta) \qquad (12)$$

The denominator of the posterior distribution (so-called marginal likelihood) is always positive and does not depend on $\theta$ and therefore plays no role in the optimization. The MAP estimate of $\theta$ coincides with the ML estimate when the prior $g$ is uniform (that is, a constant function).

In many types of models, such as mixture models, the posterior may be *multi-modal*. In such a case, the usual MAP-based recommendation is that one should choose the highest mode. However, this is not always feasible as the highest mode may be uncharacteristic of the majority of the posterior. Computation of MAP estimates can be done in several ways: (i) analytically, if the mode(s) of the posterior distribution can be expressed in closed form; (ii) numerical optimization, by using conjugate gradient method or Newton's method – this usually requires first or second derivatives, which have to be evaluated analytically or numerically; (iii) EM, which does not require derivatives of the posterior density; and (iv) Monte Carlo method, which uses simulated annealing.

## 5.3 Truth Discovery in Mobile Crowdsensing

Over the last five years, a plethora of work has been devoted to develop truth discovery algorithms for mobile crowdsensing. Figure 3 shows a taxonomy of truth discovery techniques.

The majority of the existing works have leveraged MLE-based Expectation Maximization (EM) algorithms to estimate source reliability and to infer truth value from it. Few works have also used unsupervised models based on maximum-a-posteriori (MAP), which is closely related to MLE, but employs an augmented optimization objective which incorporates a prior distribution over the quantity to be estimated. This prior is the additional information available for the entity under consideration. Thus, MAP estimation can be seen as a regularization of ML estimation. In addition, some optimization-based approaches have also been presented, which find source reliability and entity truthfulness by assuming them to be constant in alternative iterations. Privacy-preserving truth discovery and predicting missing sensed information using matrix factorization are the other miscellaneous approaches available in the literature. In the following subsections, we will give a brief introduction to these methodologies and summarize the contributions made in these works.

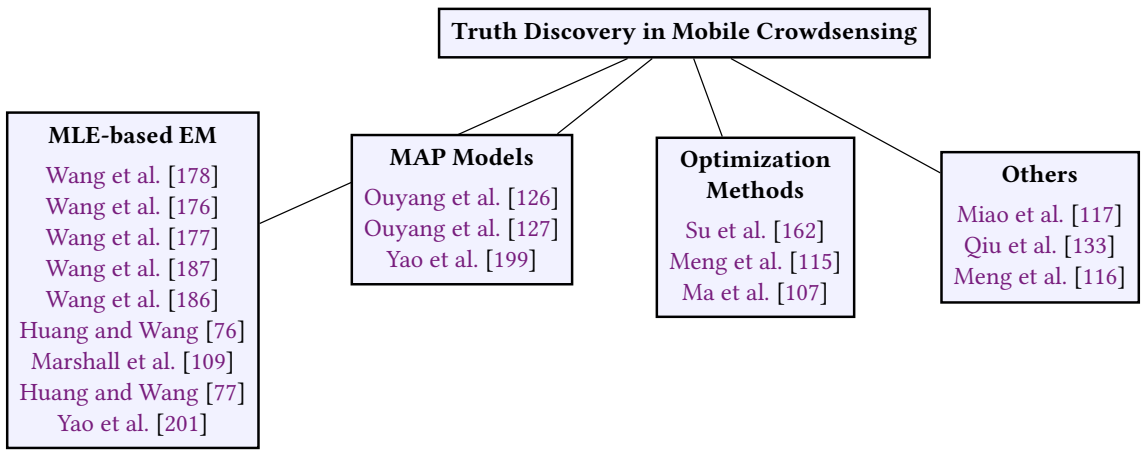

Fig. 3. Taxonomy of Truth Discovery in Mobile Crowdsensing.

*5.3.1 Maximum Likelihood Estimation-based Expectation Maximization.* The authors in [178] are the first to show that an EM formulation is appropriate to solve the truth discovery problem in mobile crowdsensing. The optimal solution directly leads to an accurate quantification of measurement correctness as well as participant reliability. This work assumes that the sources are independent and there exists no correlation among the entities. Other works use the basic EM framework and integrate different aspects of social sensing paradigm, i.e., dependency among sources [177, 201], physical constraint among sources and correlation among observed variables [176, 186], different error distribution [187] and topic expertise of sources [77].

One important aspect of EM algorithm-based truth discovery is to ensure convergence. Moreover, as mobile crowdsensing is a real-time paradigm, it becomes necessary to perform truth estimation in a reasonable time. Hence, test of convergence and computational efficiency are the two key performance indicators for any EM algorithm and we have investigated these features in the existing literature. A comparative analysis of different works which have proposed MLE-based EM algorithms for source reliability and truth inferring are given in Table 1.

As Table 1 shows, most of existing work has not given proof of convergence or analyzed the computational complexity of the proposed EM algorithm. Convergence in terms of error in estimate and corresponding iteration numbers has been analyzed in [77, 109, 178], where the EM algorithm attains a stable error rate after 3-6 rounds of iterations. Remarkably, these works have considered binary claim in terms of true and false, exception being [187] in which *uncertain* claim has also been considered. Thus, for $n$ sources, there can be $2^n$ input claim patterns from which the EM has to infer reliability and truth value. For a large number of sources, which is expected in a mobile crowdsensing application, the EM algorithm has to find optimal solution from an exponential search space, yielding non-trivial computation time in the worst case. However, these papers have not discussed any approximation to generate sub-optimal solutions in reasonable time.

*5.3.2 Maximum A Posteriori (MAP) Estimation.* The MAP estimator has been extensively used in [126, 127]. These works tackle the problem of truth estimation in quantitative crowdsourcing tasks: *percentage annotation* and *object counting* (e.g., people# in an image). The authors identified the difference of these tasks from normal object classification, and then proposed an unsupervised model, called *Quantitative Truth Finders (QTF)*. The QTF has two variants: (i) *QTF-A (annotation)*, and (ii) *QTF-M (counting)*, which model participants' bias and confidence as *Gaussian* and *Lognormal* distribution, respectively. The MAP estimation has been used to impose an *informative* prior distribution on true values and biases to restrict generation of infinite number of optimal solutions using the MLE.

| Paper | Summary | Variable | Approach | Convergence Proof |
|---|---|---|---|---|
| Wang et al. [178] | Inferring source reliability and true observation value | Binary | EM algorithm | Empirical analysis |
| Wang et al. [176] | Exploits physical (location) constraints among sources and correlation among observed variables for better estimation of reliability and truth value | Binary | EM algorithms (OtO + DV) | Not provided |
| Wang et al. [177] | Inferring correctness of individual observation and reliability of sources considering dependencies among the observations. | Binary | EM algorithm | Not provided |
| Wang et al. [187] | Truth estimator for cyber-physical systems considering different error distributions for sources | Ternary | EM algorithm (EM-VTC) | Not provided |
| Wang et al. [186] | Finds truth value and source reliability by considering correlation among the entities. Constructed the correlation by modeling the variables as Bayesian network (BN). | Binary | EM algorithm (EM-CAT) | Not provided |
| Huang and Wang [76] | Addresses the problem of determining the time (of the claim) and location (of the source) information towards truth discovery and source reliability | Binary | EM algorithm (ST-EM) | Not provided |
| Marshall et al. [109] | Assigns true values to claim and reliability to sources by incorporating the context and challenges encountered while reporting | Binary | EM algorithm (HA-EM) | Empirical analysis |
| Huang and Wang [77] | Inferred truth values of a claim by considering: (i) topic relevance and truthfulness and (ii) topic awareness and reliability of the source | Binary | EM algorithm (TA-EM) | Empirical analysis |
| Yao et al. [201] | Estimates the veracity of information shared by different sources over the social sensing platforms (e.g., Twitter, etc) by considering source dependency. This is motivated from the observation that sources tend to repeat claims of another. | Binary | Gibbs sampling has been used to approximate the error bound in situation when the number of observation patterns become intractable. | Not provided |

Table 1. Summary of Work using MLE

Truth discovery from noisy and unreliable source claims may be attempted by modeling the popularity of an event location and personal location visit tendencies. For this purpose, the authors of [127] proposed two unsupervised models, called *Truth Finder for Spatial Events (TSE)* and *Personalized Truth Finder for Spatial Events (PTSE)*, respectively. The novelty is that it preserves participation privacy by avoiding location tracking and enhances estimation accuracy by not ignoring missing reports as negative ones. Both models use the MAP estimator to find the optimal configuration of the random variables that maximize the posterior probability. Yao *et al.* [199] identifies the inefficiency related to batch analysis of offline datasets and proposes an online recursive state estimator to recover ground truth from the streaming data. The sensing time is divided into equal-sized windows and posterior belief on source is computed at the end of each window by using the prior belief obtained at the beginning through MAP estimation. A comparative analysis of different works which have used this MAP estimate to infer source reliability and truth values are given in Table 2.

*5.3.3 Optimization-based methods.* In truth discovery, the objective is to minimize the "distance" between ground truth and the inferred true values over all entities. The distance between the ground truth and any source observation is weighted by the latter's degree of reliability. This implies that if a reliable source makes incorrect claim, the truth discovery system will be more penalized while the penalty for a less reliable source will be low. A general form of the optimization problem can be formulated as follows:

$$\operatorname*{argmin}_{\{w_i\},\{x_j^*\}} \sum_{i \in \mathcal{S}} \sum_{j \in \mathcal{E}} w_i . \delta(x_{i,j} - x_j^*) \tag{13}$$

where $\delta(.)$ is the distance function that measures the difference between the information provided by source $i$, called $x_{i,j}$, and the identified truth, called $x_j^*$. Usual distance functions are: (i) 0-1 loss function for categorical data, and (ii) L2-norm can be adopted for continuous data.

| Paper | Summary | Variable | Approach | Comp. Efficiency |
|---|---|---|---|---|
| Ouyang et al. [126] | Truth estimation in quantitative crowdsourcing tasks: percentage annotation (e.g., occupancy %), object counting (people# in an image). Rather than exact match with the ground truth, this work jointly assesses latent quantitative truths and abilities of participants from noisy crowdsourced claims without any supervision. | Continuous | MAP-based unsupervised models: *QTF-A (for annotation)*, and *QTF-M (for counting)*. The QTF-A and QTF-M model participant's bias as Gaussian and Log-normal, respectively. | Asymptotic upper bound not provided. However, empirically it has been shown that QTF-A and QTF-M take 1.5 seconds to process 1000 odd tasks. |
| Ouyang et al. [127] | Addresses the problem of uncertainties with participant mobility and reliability in truth finding by avoiding location tracking and ignoring missing reports. | Binary | MAP-based unsupervised models: *Truth finder for Spatial Events (TSE)*, *Personalized Truth finder for Spatial Events (PTSE)*. | Both TSE and PTSE have been experimentally shown to take higher order polynomial time w.r.t inputs: (i) number of participants and (ii) number of events. |
| Yao et al. [199] | Proposes an online recursive state estimator for recovering ground truth from streaming data. The objective is to speed up batch-like processing of sensory data generated by social sensing platforms. | Binary | MAP-based EM algorithm which runs at the end of each time window. It combines prior belief with new samples to generate the maximum likelihood posterior belief and assertion correctness. For parameter estimation inside the time window, an interpolative estimator is invoked. | Empirically the algorithm has been shown to grow sub-linearly with the number of claims generated. |

Table 2. Summary of Work using MAP Estimate

From Equation 13 we can infer the optimization problem has two variables, namely (i) source reliability $w_i$ and (ii) identified truth value $x_j^*$. Both of them contribute to the estimation of each other in subsequent iterations. The general approach to solve such optimization problem is the *block coordinate descent* method [15], where for a particular iteration, one of the variables is assumed constant. The optimal value for the other variable is computed by setting the partial derivative of loss/error function with respect to the former to zero. Similarly, in the next iteration, the optimal value of the first variable is obtained by considering the second variable as constant, and the process continues until a predefined convergence criteria is satisfied. In the context of truth discovery, this leads to solutions in which the truth computation step and source weight estimation step are iteratively conducted until convergence.

A *Generalized Decision Aggregation (GDA)* framework has been proposed in [162] to integrate multi-modal information from distributed sensors. The framework constructs a sensor-entity *belief graph* which represents different observations. The observations for a particular entity may be categorical and source's confidence on each such category is designed in the form of a decision vector. Finally, an optimization problem is formulated which outputs the category that yields minimum error. In [115], *ubiquitous correlation* among observed entities has been considered while inferring the true value. This work developed an optimization function which minimizes two terms (balanced by a hyper-parameter): (i) the disagreement between observations and truths weighed by source reliability, and (ii) the distance between correlated entities weighed by a similarity function.

Ma *et al.* identified in [107] a limitation with existing MLE-based truth discovery approaches, and proposed a *Fine Grained Truth Discovery model for crowdsourced data (FaitCrowd)*, which assigns topics to questions, estimate topic-specific expertise of each source, and learn the true answers simultaneously. It proposed a joint optimization function which attempts to maximize two aspects: (i) likelihood of generating question content, and (ii) likelihood of generating answers. A hybrid inference approach, *Gibbs-EM*, was adopted for sampling (learn hidden variables) and variational optimization (inference). A comparative analysis of different works which have used optimization approach to infer source reliability and truth values are given in Table 3.

| Paper | Summary | Variable | Approach | Comp. Efficiency |
|---|---|---|---|---|
| Su et al. [162] | Infers truth value by incorporating source confidence | Continuous | Block coordinate descent method to solve a multi-objective (reliability, truth value) optimization | $O(mnt)$, where $n$ = number of sources (sensors), $t$ = number of events (observations), and $m$ = number of classes for each event. |
| Meng et al. [115] | Optimizes disagreement between claim and unknown truth, weighted by source's reliability, considering ubiquitous (spatial + temporal) correlation among entities. | Continuous | Block coordinate descent method to solve the multi-objective (reliability, truth value) optimization. A parallelized version of the algorithm is implemented using the MapReduce Framework. | Use of MapReduce framework on Hadoop cluster gives linear time complexity w.r.t the number of observed entities. |
| Ma et al. [107] | Estimates true value of observed variables by incorporating topic-specific expertise of each source | Categorical | Hybrid inference method combining sampling and variational optimization, named *Gibbs-EM* which is an inference method alternating between *Gibbs sampling* and *gradient descent* | The computational complexity is linear to the number of observations. |

Table 3. Summary of Work using Optimization Method

*5.3.4 Other truth discovery methods.* Few works in the literature have identified the privacy problem in the truth discovery mechanism. As sources are submitting their observations to a server in an unprotected manner, it can lead to location tracking and disclosure of other personalized information. Moreover, the reliability scores of the sources are also available to the sensing platform. In [117], a cloud-enabled *privacy preserving truth discovery (PPTD)* framework has been proposed. It sends encrypted distance value between user observation and estimated aggregated values to the cloud server and uses *homomorphic crypto-system* to update user's weight (reliability) without decrypting the received distance information. On the other hand, the user calculates the cipher-text of weighted data using the encrypted weight received in the previous step. Finally, the cloud server estimates truth based on the cipher-texts received from the users. The work in [133] is different from standard truth discovery, however, the broad objective to improve the quality of information in social sensing paradigm remained the same. The authors addressed the challenges of privacy and multimedia data quality preservations in mobile crowdsensing applications and proposed *SLICER*, a *k*-anonymous privacy preserving scheme. It integrates *erasure* data encoding scheme with message exchanging and ensures that the service provider cannot identify the contributor of each sensing record from a group of at least *k* participants.

In [116], the authors identified both *sparseness* and *redundancy* in crowdsensing data which affects the overall truth estimation accuracy. Interpolation infers true values of missing entities based on entity similarities, however, it considers redundancy and sparsity as separate challenges. The authors proposed *Redundancy and Sparsity Tackling (RST)* framework which designs an optimization problem to minimize error in truth inferring. This optimization integrates three terms in a weighted (given by hyper-parameters) sum form:

- *Matrix factorization term (MF)*: It identifies $K$ virtual users such that two latent matrices $U \in \mathbb{R}^{M \times K}$ and $V \in \mathbb{R}^{N \times K}$ are formed from a big source-claim matrix of the order $M \times N$, where $M$ and $N$ are the numbers of source and entity, respectively. Next, these matrices are used to provide a good approximation for each entry of the source-claim matrix $X \approx UV^T$.
- *Regularization term on entity similarity (R1)*: This term is used to extrapolate the values for some entities which have not observed by any virtual user. It essentially incorporates a similarity relationship based on a basic principle - observations from virtual users should not differ too much on entities that are similar to each other.
- *Regularization term on virtual users (R2)*: By this regularization, the framework assigns *importance* to virtual users. If a virtual user is more important, higher penalty will be received if its observation is quite different from the aggregated one.

Table 4 presents a comparative analysis of different works which have used miscellaneous approaches to infer user reliability and truth values.

| Paper | Summary | Variable | Approach | Convergence Proof | Comp. Efficiency |
|---|---|---|---|---|---|
| Miao et al. [117] | Performs truth discovery by ensuring privacy of the user's data and reliability score using a crypto-system. Parallel PPTD scheme implemented using the MapReduce Framework. | Categorial, Continuous | Privacy-preserving truth discovery (PPTD) iteratively updates reliability and truth value through an iterative approach. Information exchange between two phases is protected by a cloud-based homomorphic crypto-system. | Empirical analysis | Truth estimation takes much more time than weight update. However, total computation time is linear to the number of observed entities |
| Qiu et al. [133] | Addresses the challenges of privacy and multimedia data quality preservations in mobile crowdsensing applications. | Continuous as records | *SLICER*: Erasure coding-based algorithm (for data quality) + two transfer schemes TMU and MCT (for preserving privacy). | Erasure-coding: deterministic algorithm; transfer schemes have been reduced to *0-1 knapsack problem* for NP-completeness. Greedy approach proposed for sub-optimality. | Computational complexity is linear to the sizes of the sensing records. |
| Meng et al. [116] | Estimates the true values of the entities from redundant and sparse data in crowd sensing applications. | Continuous | Developed Redundancy and Sparsity Tackling (RST) framework. It uses matrix factorization, regularizations on entity similarity and virtual users to predict missing values and finally aggregate them to infer the truth. Block coordinate descent method is used to find out the optimal values of each parameter. | Theoretical analysis | Not provided. |

Table 4. Summary of Work using Miscellaneous Approaches

## 5.4 How to Improve Existing Work

In this section, we have analyzed the state-of-the-art research on the topic of truth discovery, which is by far one of the most promising approaches for QoI estimation and enforcement in mobile crowdsensing. Truth discovery approaches can be divided into four broad categories: (i) MLE-based

EM algorithms, (ii) MAP-based unsupervised models, (iii) Optimization-based methods, and (iv) a few assorted approaches related to privacy-preserving and guessing the missing information.

Several limitations still linger in existing work, which we expect will be addressed in the near future, and are enumerated below.

(1) Most of the existing approaches are iterative and computationally expensive. Moreover, convergence guarantees have also not been provided in majority of the EM algorithm-based schemes. As crowdsensing is real-time application, there is a need to propose lightweight QoI estimation mechanism;
(2) It is still unclear how to address the problem of malicious participants who may collude to generate a false consensus regarding an event. Further, reliable users may also be compromised by an adversary and generate false observations. In such cases, mere consideration of source weights will not help in inferring correct truths. Most of the state-of-the-art papers have ignored this aspect and assumed ideal attack-free sensing environment;
(3) Reliability of sources has been considered to be static. This again undermines the fact that the sensors in this paradigm are humans and not dumb nodes. As human does not always behave rationally, a better model is required to capture dynamism in the reliability scores;
(4) Most of the works have considered *binary* claims to realize their models in a simplistic manner. However, in reality, crowdsensing data can be multi-model viz., text, audio clip, video, image, and so on. Truth discovery on such data is non-trivial and brings new challenges;
(5) Crowdsensing is an "open" system, where any user can get himself registered and participate in the sensing task. There exists no bounds either on the number of users or the observed entities in the system. Thus, scalability of the truth finding mechanism is an important requirement. By scalability, we imply that truth can be inferred in reasonable time event if the number of participants and the entities increase. Very few works in the existing literature have considered this issue;
(6) With the exception of [200], EM algorithm-based works are mostly designed to run on static datasets. However, the computation becomes less efficient for streaming data as the iterative algorithm needs to re-run on the whole dataset from scratch every time it gets updated. Additional research should address this important issue.

## 6 TRUST FRAMEWORKS

In this section, we introduce and discuss in details existing trust frameworks for mobile crowdsensing. First, in Section 6.1 we survey the foundations of trust in computer science and distributed systems, and show their analogy with QoI in mobile crowdsourcing systems. We also introduce the "trust loop" problem in mobile crowdsensing and briefly discuss related work. Then, by referring to existing work, we discuss the methods and approaches for collection of trustworthiness evidence in Section 6.2 and computation of trust scores in Section 6.3. We conclude the section by discussing how to improve existing work in Section 6.4.

### 6.1 Overview, Problem Definition, and Related Work

The original concept of trustworthiness has its origins in human society. According to the Merriam-Webster dictionary [195], trust is the *"[...] belief that someone or something is reliable, good, honest, effective, etc."* In other words, trust is a qualitative way of expressing whether a particular interaction with an entity is going to be dependable, given evidence of outcomes of prior or current behaviors and interactions. Over the years, the concept of trust has significantly evolved and has also been employed in the computing, communication and networking fields. Analogically, each node in a network can be viewed as a human, whereas the network itself may be seen as a society where decisions are made based on inputs/contributions from each node. This idea is often termed as mechanical or *technological trust*, a quantitative measure of trustworthiness for interactions between entities [67]. In the context of mobile crowdsensing, services are offered based on the contributions/information provided by the participating users, either through mobile devices or automated sensors [92]. Hence, it is reasonable to relate the trustworthiness of the overall contributions from multiple mobiles and sensors to the Quality of Information (QoI).

In order to represent, quantify, and manipulate trust between different entities, several metrics have been introduced. More formally, if $\mathcal{E}$ defines the set of considered entities, also called as *trustees*, the *trust function* may be defined as $\mathcal{T} : \mathcal{E} \rightarrow \mathcal{D}$ that associates to each entity $e \in \mathcal{E}$ a trust "value" $\mathcal{T}(e)$, also referred to as "score", where $\mathcal{D}$ is the *domain* of the trustworthiness metric.

Several metrics have been defined to quantify trustworthiness, which are briefly surveyed below. For a comprehensive survey of the topic, the reader may refer to [16, 152].

- *Single-valued.* This metric assigns a single trust value in a domain $\mathcal{D} \subseteq \mathbb{R}^+$. Generally, the lower the value, the less trustworthy an entity is. The values in $\mathcal{D}$ are usually bounded between +100 and −100 [63], 0 and 1 [120], or −1 and +1 [40, 108], and may be discrete or continuous. Often, single-valued metrics are inspired from Bayesian statistics [14], and represent trustworthiness as the probability that a trustee will perform an action that is beneficial or at least not detrimental to the trustor. In this case, the metric also observes the *Cromwell's rule* [80], which prohibits the assignment of 0 or 1 as trust because if the prior probability is 0 or 1, then according to Bayes' theorem, the posterior probability is forced to be 0 or 1 as well. In this case, no evidence, however strong, could have any influence on the posterior, which may be improper for a system where Bayesian inference is used based on incremental evidence over time.

- *Multi-valued.* This representation is the most popular when the system is interested in modeling and characterizing the uncertainty of trustworthiness. This is achieved by characterizing trust not only by a single value in $\mathcal{D}$ but instead through a set of values. For example, in some cases, trustworthiness is defined by a fuzzy set in a continuous [26] or discrete [135] space. Thus, the trustworthiness is represented by the shape of the fuzzy set – a wider fuzzy set means less confidence on that reputation, and vice versa. These fuzzy sets are then manipulated and updated through fuzzy aggregation [94]. Another very popular model for multi-valued representation is the Jøsang's belief model [87] which uses a quadruplet termed as *opinion* to express all components about the interaction with an agent. More formally, an opinion about entity $e$ is represented as a 4-tuple $\omega_e = (b, d, u, a)$, where the components represent degrees of belief, disbelief, uncertainty, and relative atomicity respectively, where $\{b, d, u\} \in [0, 1]$ and $b + d + u = 1$. The expected opinion $\mathcal{T}(e) = \mathbb{E}(\omega_e)$, is computed as $\mathbb{E}(\omega_e) = b + a \cdot u$. The relative atomicity $a \in [0, 1]$ determines the extent to which the degree of uncertainty contributes to $\mathbb{E}(\omega_e)$. For example, if we are trying to find whether a node in a network is malicious, then $a = 0.5$ assuming no other information is available on that node.

Another concept that arises when dealing with trust is *reputation*. Although very similar in purpose, the concepts are distinct. Specifically, conversely from trust, reputation does not necessarily involve two parties only and can represent a multinomial nature of interactions with multiple parties. Therefore, the term reputation can be loosely defined as, "*what is generally said or believed about a person's or entity's character or standing* [90]. In other words, it is a combined measure of trustworthiness as seen by all other entities interacting with a particular trustee. Since the interaction between the mobile crowdsensing system and the participating users can be seen as to one-to-one interaction (i.e., participants are unaware of each other), for the sake of simplicity we will use the terms "reputation" and "trust" interchangeably, as well as "system" and "framework".

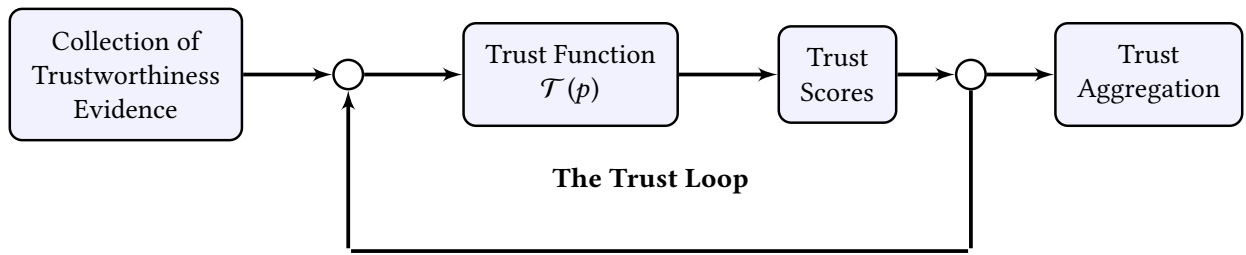

Fig. 4. Block scheme of interactions in the trust loop.

*6.1.1 Problem definition.* Similarly to truth discovery, the main purpose of existing trust-based frameworks is to assess the QoI of the data submitted by a participant $p$. This is done by inferring the reliability of $p$ through the computation of his/her trustworthiness. There are several activities and algorithms involved in such process, whose interactions are defined by us as "the trust loop". Figure 4 depicts a block scheme of the activities and interactions involved in the trust loop.

The computation of trust values for each participant $p$ is done through the trust function $\mathcal{T}(p)$, which takes as input (i) the trust value of $p$ computed at the previous time step (if any); and (ii) the evidence towards the trustworthiness of $p$ collected during the current time step. The trust function then computes the current trust score of participant $p$, which is then used to infer the QoI of the data submitted by $p$ during the current time step. This is usually done through aggregation of trust scores of participants reporting on the same phenomenon.

*6.1.2 Related work.* Beyond mobile crowdsensing, the notion of *computational trust* has been employed to solve many research problems related to the QoI, including e-commerce [93, 124]. Among the fields closely related to mobile crowdsensing, we include sensor networks [9, 83, 84, 137, 163], opportunistic networks [32, 99, 153, 198], and social networks [59, 197], to name a few. However, establishing trust in mobile crowdsensing poses some unique challenges that cannot be addressed solely with existing trust frameworks. These challenges include (i) the need of scalability given by the sheer scale of mobile crowdsensing systems; (ii) the need to collect trustworthiness evidence from likely unreliable sources; and (iii) the need to handle sensing reports of different types, including multimedia information (e.g., photo or video).

## 6.2 Collection of Trustworthiness Evidence

Earlier, we have defined trust as the likelihood that a participant will submit reliable information, based on the past behavior of the participant. Thus, regardless of the trust function being used by the system, it is mandatory to collect objective measurements (i.e., *evidence*) that may hint at the participants' degree of reliability.

Arguably, the collection of trustworthiness evidence is the most challenging activity in the trust loop. Specifically, even if some evidence of reliability is obtained, how can we trust that this very evidence will itself be reliable? How can we scalably collect such information given the large scale of existing mobile crowdsensing systems? In order to address these challenges, several approaches have been proposed over the last few years. Below we summarize the most relevant ones, and also discuss the advantages and disadvantages of each approach.

- *Use of Contextual data.* The majority of existing work has focused on the use of contextual information to collect evidence towards establishing trustworthiness of participants. By contextual, we mean information that (i) is related to the particular event being monitored; and (ii) indicates the suitability of each participant to monitor that event. In most cases, collecting contextual data does not require additional equipment and can be easily implemented, which is a major advantage. Unfortunately, some forms of contextual data (e.g., spatio-temporal, personal endorsement) can be forged by malicious users interested in increasing their reputation and thus obtain some form of gain, e.g., monetary rewards [141, 144, 147].

(1) *Spatio-temporal location.* Several works [140, 189, 191, 193, 211] have proposed the use of spatio-temporal location of sensing reports to collect trustworthiness evidence. The rationale is that if the report is originated from a location closer to the event being monitored, and has been submitted before a given deadline, then the report and thus the participant is trustworthy. Some works have also considered frequency of submissions [144] and the timeliness, which measures how promptly a participant performs prescribed tasks [5, 140]. For example, in [189] the authors propose a spatio-temporal provenance factor to establish trustworthiness. Due to privacy reasons, the server does not know the actual location of the participant. Instead, a circular *cloaking area* of diameter $D_c$ is reported as the participant's location. The server assigns to the center $L_s$ of this cloaking area the approximate *sensing location*. The task of interest has a location $L_t$ tagged as the *target location*. Thus, the closer $L_s$ is to $L_t$, the more trustworthy the report is. More formally, the authors in [189] quantify a *location distance factor* as

$$\theta = e^{-D_c \alpha} \cdot \left(1 - e^{-(|L_s - L_t|)\alpha}\right) \quad \text{for } \theta \in [0, 1) \tag{14}$$

where $\alpha$ is the known as the *location sensitivity parameter* that controls the weight of the location factor's influence. If $|L_s - L_t|$ is larger than a maximum threshold the location is considered inappropriate, then the reports in such cases are discarded. In [189], the authors also introduce a *time gap factor* similar to the location distance factor. Finally, the base trust of a report is a function of the location distance and time gap factors. Other contextual pieces of information, known as *milieu factors*, are also included in the computation. These milieu factors include type of data, mobile device type, speed of the mobile device, battery level, weather conditions around the expected location, and so on.

(2) *Personal endorsement.* A number of works in the literature [6, 105] propose the use of personal endorsement over social networks to measure the trustworthiness of participants. The rationale behind this approach is that economically, the participants aim to maximize their own economic benefit by (optimally) selecting whom to endorse and whom to be endorsed by. The advantage of such approach is that although a participant might generate fake or noisy data, either by sabotage or by mistake, he tends to refrain from doing so if (a) such conducts may be witnessed by other people, and/or (b) his connections have to bear the consequences, such as compromised benefit. Such a witness effect and beneficiary effect have the potential to incentivize participants to improve the quality of their contribution [106]. In [6], the authors proposed an approach based on a preferred participant list to select the most appropriate individuals to execute the sensing tasks. This list may be automatically generated by the application (typically containing the participant's friends who have demonstrated trustworthy behavior) or manually by the requester (based on his trust upon these individuals). Similar to the preferred list, a blocked list may also be available which contains the list of those whom the participants desire not to endorse.

(3) *Expertise.* The rationale of this approach is that greater trust is placed in participants having expertise in executing the task. *Expertise* is defined as the measure of a participant's knowledge and is particularly important in tasks that require specific knowledge about a particular domain, such as programming skills, familiarity to a geographical area, proficiency with a particular language, and so on. Expert finding systems such as [18, 103, 181] may be employed to evaluate expertise that analyzes explicit (e.g., public profile data and group memberships) as well as implicit information (e.g., textual posts) to extract user interests and fields of expertise [5]. In [6], the authors formalized this concept and introduced the concept of *expertise level*. They assumed an expert system in place which crawls a social network graph to identify and extract the field of expertise of various participants. More formally, the expertise level is given as $E = \eta(RE \cap PE)/\eta(RE)$, where set $RE$ denotes the set of required expertise, $PE$ denotes the set of participant's expertise, and $\eta(A)$ is the number of elements in a given set $A \in \{RE, PE\}$. This approach may be helpful when the system requires the execution of a very specific task that requires prior expertise, and in general may increase the overall trustworthiness. However, the drawbacks are that this approach does not take into account malicious behavior, and that expert recruitment systems may be impractical in large-scale mobile crowdsensing systems.

- *Use of Outlier Detection and Similarity.* Some existing works [79, 205] have proposed the use of outlier detection techniques [48] to obtain trustworthiness evidence. This approach is rooted in the idea that in a typical mobile crowdsensing application, the system often has access to multiple data that characterize the same physical phenomenon but are sent by different participants. Given such data redundancy, performing outlier detection may help inferring the trustworthiness of participants. Outlier detection determines the likelihood of a data coming from an untrusted source by measuring its distance to a common value (e.g., average fuel price); the smaller the distance, the more trustworthy is the device. A drawback of this approach is that it considers only the information in the current observation period while ignoring the historical information.

  Outlier detection can be broadly classified as either model-based or consensus-based techniques [73]. A model-based approach requires a priori knowledge of the underlying physical process in which the application is interested, which may be difficult to acquire in most contexts. On the other hand, consensus-based techniques work on detecting inconsistencies among data, and use the deviations from a common consensus to identify outliers [161]. This approach may work in most cases, however it does not take into account collusion-based attacks by malicious users [7, 180, 191].

  By following the same idea, the concept of similarity has also been leveraged recently [140, 192]. In particular, for two sensing reports $i$ and $i'$, the similarity score $S(i, i')$ between the two reports is a quantity between $-1$ and $1$ where $-1$ means completely conflicting with each other and $1$ means exactly the same. In [192], the trust score takes into account the similarity score, by considering a similarity factor $\Delta_i$ to information item $i$ belonging to a collection $C_i$ of reports as follows:

  $$\Delta_i = \frac{\sum_{i, i' \in C_i, i = i'} S(i, i')}{|C_i| - 1} \quad (15)$$

  where $|C_i|$ is the number of reports in the collection. Each report in a collection will be assigned with a similarity factor. A negative similarity factor means there are more conflicts in the collection while a positive similarity factor means there are more supports.

- *Ground-truth data.* The use of limited ground-truth data to acquire trustworthiness evidence has been employed in prior work to ease the impact of malicious participants. In [141, 144], we have proposed to utilize a limited number of *mobile trusted participants* (MTPs) to help build reputation scores in a secure manner, and thus 'bootstrap' the trust in the system [27]. Specifically, MTPs are participants hired by the sensing application to periodically generate reliable reports that reflect the actual status of the event being monitored around their locations. A similar methodology is being successfully used in the *National Map Corps* project [112] developed by the U.S. Department of Geographical Survey (USGS), where USGS employees are recruited to validate the crowdsourced data, such as the exact location of schools and cemeteries. Furthermore, in the *Crowd Sourcing Rangeland Conditions* project [82], Kenyan pastoralists are recruited by researchers to validate the sensed data regarding local vegetation conditions. A similar concept has been explored in [131], where the authors use *anchor nodes* (i.e., participants completely trustworthy) to improve reputation scores.

We point out some advantages of using MTPs to collect trustworthiness evidence: **(i)** the approach does not assume any particular mobility pattern, speed or trajectory of the MTPs; and **(ii)** conversely from endorsement- or spatio-temporal location-based approaches, MTPs guarantee trustworthiness through the evaluation of the accuracy of the *data* provided by the participant. This gives the approach resiliency to GPS-spoofing-based attacks, since the trustworthiness of sensed data will be guaranteed *regardless* of the accuracy of the participants' locations. In addition, the use of MTPs may help prevent collusion-based attacks by users. The downside of MTPs is that they also inevitably represent an additional *cost* for the mobile crowdsensing system, as MTPs need to be recruited.

The use of low-level sensor data has also been proposed to collect evidence of trustworthiness. In particular, GPS, accelerometer, gyroscope, compass, microphone, etc. can be used to verify whether the participant's voluntary high-level input is consistent with the observations from low-level sensor data. For example, let us suppose a participant submits a report indicating a heavy traffic jam. This report is not going to be trustworthy if the device accelerometer shows that it is in high-speed motion. Therefore, this approach looks for a consistency between the participant's voluntary input with the involuntary opportunistic input. Ad-hoc connectivity of smartphone devices through WiFi and Bluetooth may also be used to validate the smartphone's position and avoid malicious behavior, as proposed in [165, 180, 193]. The disadvantage of this approach is that an adversary may still forge low-level sensor data, unless modules such as the ones in [156] are used to verify sensor data.

### 6.3 Computation of Trust Scores

There exist been several approaches in the literature as regards to how to compute trust scores for mobile crowdsensing participants. In the following, we summarize the most relevant ones and discuss advantages and disadvantages.

*6.3.1 Gompertz's Function Model.* Gompert'z function [62] is a mathematical model for a time series, where growth is slowest at the start and end of a time period. The right-hand or future value asymptote of the function is approached much more gradually by the curve than the left-hand or lower valued asymptote, in contrast to the simple logistic function in which both asymptotes are approached by the curve symmetrically. It is a special case of the generalized logistic function. The equation is defined as $G(t; a, b, c) = a \cdot e^{b \cdot e^{c \cdot t}}$, where $a$, $b$ and $c$ are the distribution's parameters. Figure 5 shows the Gompert'z function with different $b$ and $c$ values. It is traditionally used to model the population growth in a confined space, as birth rates first increase and then slow as resource limits are reached [25, 212].

Huang *et al.* proposed in [78, 79] a trust framework exemplified in the context of a urban noise monitoring system, where it is assumed the adversary compromises the hardware to generate spurious sensor readings. Every participating device's sensing report is labeled as cooperative or non cooperative based on the output of an outlier detection algorithm. The proposed reputation system works over such a binary evidence. The system updates the trust scores every time period of duration $T$, where each period is labeled as $k$. An *epoch* contains a number of time slots denoted by $t$. Sensor readings from any device $i$, over a duration $T$ is denoted as $X_{i,k} = \{x_{i,t}, \cdots, x_{i,t+T-1}\}$.

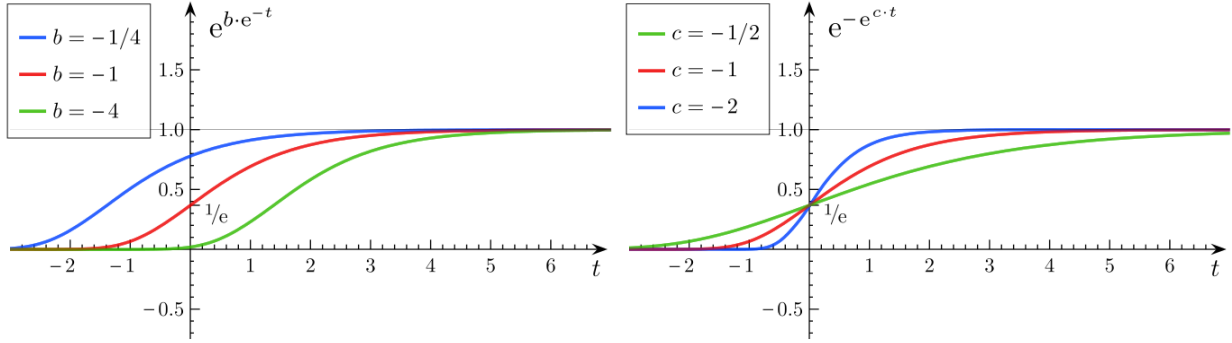

Fig. 5. Gompertz's function with different $b$ and $c$ parameters.

A "watchdog" module takes in $X_{i,k}$ as input and generates a rating level denoted by $0 < p_{i,k} < 1$, which is an average calculated based on a distance similarity measure between the advertised $x_{i,t}$ and the consensus $r_t$. All $p_{i,k}$ are initialized to $\frac{1}{n}$, and are iteratively updated in every $k$-th epoch, such that all past ratings of device $i$ over past epochs and current epoch is denoted as $p_{i,k'}, \forall k' = 1, \cdots, k$. Based on this information, the reputation of the device $i$ is calculated by using the Gompertz's function, which converts past cooperation levels into a cumulative score $R_{i,k}$. The input to the Gompertz's function $p'_{i,k}$ is given as

$$p'_{i,k} = \sum_{k'=1}^{k} \lambda^{(k-k')} p_{i,k} \quad (16)$$

where $0 < \lambda \geq 1$ is known as an aging parameter that controls how quickly past observations are forgotten. The rationale is that, if the cooperation levels are consistently high, so must be the reputation score. The reputation score is at first low and then it increases non-linearly if the participant is cooperating effectively.

The advantage of using the Gompertz's function is the ease of implementation and its efficacy in modeling trust, as it effectively captures dynamic behavior of the participants. However, as pointed out in [144], the approach does not take into account the possibility that a substantial number of participants may be malicious.

*6.3.2 Jøsang's Belief Model.* In a previous work [144], we proposed a trust model based on Jøsang's belief model [87] that combines the reliability and participation level of users to produce an aggregate trust score, which is then used to decide (i) the incentive or pay-off level of the participant, and (ii) whether or not accept his/her sensing reports. Mobile trusted participants (MTPs) are used to collect evidence of trustworthiness. More formally, at every time step $j$ and for every user $i$, the contribution is compared with that of an MTP. A 'match' event occurs if the MTP input agrees with the user $i$'s contribution. Likewise, a 'mismatch' event occurs if they do not agree. And an 'uncertain' event occurs if the MTP is not available at that time $j$. Jøsang's opinions are used to quantify both reliability and trustworthiness levels, which are updated reflecting a match or a mismatch with the MTP's contribution (if any) at a particular time step.

The final combined measure of the reliability and trustworthiness opinions (respectively $\omega_{i,j,r}$ and $\omega_{i,j,t}$) is given by the conjunction operator defined as follows:

$$\omega_{i,j,rt} = \begin{bmatrix} b_r \cdot b_t \\ d_r + d_t - d_r \cdot d_t \\ b_r \cdot u_t + b_t \cdot u_r + u_r \cdot u_t \end{bmatrix} \quad (17)$$

The reputation $r_i^j$ of user $i$ and time $j$ is computed using the expectation operator as discussed in Section 6.1. Similarly, in [141] the author models the probability of deeming a sensing report reliable in the case it has not been validated by an MTP as follows:

$$\mathbb{P}\{A \mid \overline{V}\} = \mathbb{P}\{V\} \cdot \mathbb{P}\{\overline{F}\} + \frac{1}{2} \cdot \mathbb{P}\{\overline{V}\} \tag{18}$$

where $\mathbb{P}\{\overline{F}\}$ is the probability that the participant will submit unreliable information. This formula is inspired from Jøsang's trust model and can be explained as follows. The first part, $\mathbb{P}\{V\} \cdot \mathbb{P}\{\overline{F}\}$, represents the "degree of belief" in the participant; it is higher when the user is validated most of the time ($\mathbb{P}\{V\}$ close to 1) and the reports are reliable. The second part, $1/2 \cdot \mathbb{P}\{\overline{V}\}$, represents the "degree of uncertainty" in the users; it is higher when most of the reports have not been validated. Note that, as $\mathbb{P}\{V\}$ increases, the value of $\mathbb{P}\{A \mid \overline{V}\}$ approximates to $\mathbb{P}\{\overline{F}\}$. Also, if $\mathbb{P}\{V\} = 0$ (i.e., no MTPs are present), the system deems as reliable every report with probability $1/2$ (coin tossing), as there is no reason to be more inclined to accept or reject the report if no information is available.

*6.3.3 Beta Distribution Model.* Some existing work has leveraged the Beta distribution [89] to compute trust scores [56, 138]. Usually the Beta distribution is suitable when the evidence state space is binary. However, evidence with a ternary state space can be alternatively modeled by coarsening the evidence into a binary space [88], approximating the third state space into partially positive and partially negative ratings/feedbacks/outcomes.

More formally, the Beta distribution suggests that if $r$ is the number of positive and $s$ is the number of negative outcomes from a set of $N = r + s$ number of interactions with an entity, then the trust of that entity over this action can be modeled as a Beta distribution with parameters specified by $\varphi = r + 1$ and $\kappa = s + 1$, such that the posterior probability density function given as:

$$f(p|\varphi, \kappa) = \frac{\Gamma(\varphi + \kappa)}{\Gamma(\varphi) \cdot \Gamma(\kappa)} \cdot p^{\varphi - 1} (1 - p)^{\kappa - 1} \tag{19}$$

A Bayesian estimation of parameters shows that the mean of the probability density function in Equation (19) can accurately model trust metrics given by

$$E = \frac{\varphi}{\varphi + \kappa} = \frac{r + 1}{r + s + 2} \tag{20}$$

As demonstrated in [79], the Beta trust scoring takes a less aggressive approach in penalizing users that contribute corrupted data. On the other hand, the period over which a participant may contribute corrupted data may potentially be short-lived (particularly, when this happens unintentionally). Furthermore, the frequency of occurrence of such events may also be high (e.g., a user may frequently place her phone in the bag while it is collecting noise pollution samples). In such cases, the Beta scoring may not be the most effective way to capture such dynamic behavior.

*6.3.4 Dirichlet Model.* This trust model is used when the state space of the trustworthiness evidence is not binary but instead follows a multinomial distribution [138, 196]. The multinomial distribution is the generalization of the binomial distribution when $k > 2$ potential outcomes are possible, where each trial results in one of the $k$ outcomes from a set of $N$ possible trials. Therefore, the count of observations may fit very well with this model. Specifically, observations data $D$ for any participant can be treated as a multinomial distribution, provided the probabilities of occurrence of each outcome are mathematically represented as $p(D|X)$. Here, $D$ is a vector representing the observed data, such that $D = \{d_1, d_2, d_k\}$. Similarly, $X$ represents the probability vector denoting the probability of occurrence of each outcome such that $X = \{x_1, x_2, x_3\}$. In general these are represented as $D = \{d_i | 1 \leq i \leq k\}$ and $X = \{x_i | 1 \leq i \leq k\}$.

However, initially $X$ is unknown and hence the problem is estimation of parameters in $X$ given evidence or data $D$. The probability density function (PDF) of a Dirichlet distribution returns the subjective probabilities $x_i$ (also known as degree of belief) for $K_i$ number of competing events given each possible outcome $i$ has been observed $d_i - 1$ times. Given that observation count $d_i - 1$ is known from the gathered evidence, we can calculate the probabilities of each event. More formally, the Dirichlet probability density function with variables $\vec{x}$ and parameters $\vec{d}$, is defined as

$$f(\vec{x}|\vec{d}) = \frac{\Gamma(\sum_{i=1}^{k} d_i)}{\prod_{i=1}^{k} \Gamma(d_i)} \prod_{i=1}^{k} x_i^{d_i-1}, \tag{21}$$

The relation between observation parameter $d_i$ and actually observed outcomes $r_i$ is that $r_i + C \cdot a_i = d_i$, where $\sum_{i=1}^{k} a_i = 1$ and $C > 0, a_i > 0$ such that zero occurrence of an outcome preserves the condition that $d_i > 0$.

Obtaining the trust score is equivalent to computing the expectation of the Dirichlet distribution, which is computed as

$$\mathbb{E}(x_i|\vec{d}) = \frac{d_i}{\sum_{i=1}^{k} d_i} \tag{22}$$

where $d_i$ is known as the total evidence count for event $i$. The degrees of belief associated with the outcomes are expressed as the mean of each outcome.

*6.3.5 Fuzzy Logic Model.* Prior work [3–6] has suggested to leverage fuzzy logic [42] to compute trust scores based on the collected trustworthiness evidence. In such work, it is supposed that every participant is part of an existing online social network platform. The system works in a publish/subscribe manner where a *requester* publishes a sensing task. A set of contributors from her social network graph known as *participants* are responsible to report their observations related to the sensing task. Any requester and contributor pair form a friendship relation. Every requested task is assumed to have a deadline $d$ associated with it.

The authors of [3–6] assume the trustworthiness level of a contributing user is affected by the following five factors that are a combination of personal and social factors. The personal factors include expertise (E), timeliness (T), location (L), friendship duration (F), and interaction time gap (I). The proposed trust frameworks leverage fuzzy logic to calculate a comprehensive trust rating for each contribution, referred to as the Trust of Contribution (ToC). The inputs to the fuzzy inference system are the crisp values of the Quality of Contribution (QoC) and Trust of Participant (ToP). The QoC is a value between 0 and 1 and reflects the quality of the sensing report (i.e., accuracy, precision, timeliness) and is determined by external outlier detection algorithms. On the other hand, the ToP is derived as a weighted sum of the five personal factors, where the weights control the extent to which each of these factors affect the trustworthiness.

The trust frameworks derives the ToC as a function of the ToP and QoC by using a fuzzy logic inference system. The steps involved in such computation are described as follows.

(1) *Fuzzifier.* Since the ToP and QoC are crisp values (i.e., a scalar value and not fuzzy sets), the fuzzifier converts these values into a linguistic variable. In other words, it determines the degree to which these inputs belong to each of the corresponding fuzzy sets. The fuzzy sets for QoC, ToP are defined as {Very Low, Low, Medium, High}, while the ToC fuzzy set is {Very Low, Low, Medium, High, Very High}. For any set $X$, a membership function $\mu(X) : X \rightarrow [0, 1]$ quantifies the grade of membership of an element $x \in X$ to the fuzzy set $X$. The value of 0 means that $x$ is not a member of the fuzzy set and 1 means that $x$ is fully a member of the fuzzy set. The values between 0 and 1 characterize fuzzy members, which belong to the fuzzy set only partially.

| Paper | Summary of Contributions | Evidence Collection | Trust Scoring |
|---|---|---|---|
| Amintoosi et al. [6] | Framework for identification and recruitment of well-suited participants in large-scale online social networks with unknown network topology and complex friendship relations. | Spatio-temporal Location, Endorsement, Expertise | Fuzzy logic |
| Amintoosi and Kanhere [5] | Application-agnostic reputation framework with utilization of PageRank algorithm as the basis of trust. | Spatio-temporal Location, Personal Endorsement, Expertise | Fuzzy logic |
| Huang et al. [78, 79] | Proposes for the first time the use of Gompertz's function for trust scoring. | Outlier Detection | Gompertz's Function |
| Luo et al. [106] | Introduces a social concept called nepotism by linking mobile users into a social network with endorsement relations, and proposing the overlaying of this network with investment-like economic implications. | Endorsement | Customized |
| Ren et al. [140] | Proposes a participant selection scheme to choose the well-suited participants for the sensing task under a fixed task budget. | Spatio-temporal Location, Endorsement, Similarity | Customized |
| Wang et al. [188, 189] | Derives a trust model based on information similarity and path difference, present a mechanism to detect and reduce the effect of collusion attacks. | Spatio-temporal Location, Similarity | Customized |
| Wang et al. [191] | Discusses a trust model based on information similarity and path difference, present a mechanism to detect and reduce the effect of collusion attacks. | Spatio-temporal Location, Similarity | Customized |
| Yu et al. [205] | Introduced the accumulated reputation model (ARM), where trust is computed and accumulated based on sensing data. | Outlier Detection | Gompertz's Function |
| Restuccia [141, 144] | Proposed trust framework based on Mobile Trusted Participants (MTPs). | Outlier Detection | Ground Truth, Jøsang's model |

Table 5. Summary of Trust Frameworks for Mobile Crowdsensing

(2) *Inference Engine.* The role of inference engine is to convert fuzzy inputs (QoC and ToP) to the fuzzy output (ToC) by leveraging If-Then type fuzzy rules. To define the output zone, the max-min composition is used [94].
(3) *Defuzzifier.* A defuzzifier converts the ToC fuzzy value to a crisp value in the range of [0, 1]. The Centre of Gravity (COG) [172] defuzzification method is perhaps the most commonly used and popular defuzzification technique with the advantage of quick and highly accurate computations.

### 6.4 How to Improve Existing Work

In this section, we have surveyed relevant existing work towards the development of effective and efficient trust frameworks for mobile crowdsensing. Table 5 summarizes the most relevant contributions discussed in this section. Trust frameworks have garnered significant attention from the research community, given they possess many desirable characteristics that make them an effective choice for implementing QoI in mobile crowdsensing. Among them, (i) their ease of implementation; and (ii) their flexibility and capability to adapt to dynamic behavior of participants.

Despite the advantages provided by the trust framework, several research challenges still linger and need to be addressed by future work, which are enumerated below.

(1) The approach of contextual information for trustworthiness evidence suffers from many disadvantages. Most importantly, malicious participants can tamper with their smart devices or use apps that can forge contextual information, rendering the trust assessment process useless [193]. Additional research must be devoted on finding the right scenarios where it is appropriate to use contextual information to assess trustworthiness.

(2) The impact of different participants' behaviors on the trust scores of participants is still unclear. In particular, there is a need to *mathematically model* the participants' behavior and predict the optimum parameters that may not only make existing trust frameworks 100% robust against carefully planned attacks, but also stimulate good behavior among users. Although there exists some research work in this direction [38, 167, 169, 207], the literature still lacks a comprehensive investigation on these topics.

(3) Most of existing work have not investigated the scalability and computational complexity of known trust frameworks. This aspect is fundamental given the large scale of mobile crowdsensing systems. Rigorous theoretical analysis, as well as implementation on parallel frameworks such as *MapReduce* [35], are necessary to address the scalability issue and ensure that proposed trust frameworks are implementable in real-world scenarios.

## 7 OPEN RESEARCH CHALLENGES AND DIRECTIONS

In this section, we propose novel research challenges pertaining to QoI in mobile crowdsensing.

### 7.0.1 Assessing Trustworthiness of Complex Information.
We point out that most of existing works have considered binary (i.e., true/false) or multinomial (i.e., belonging to a set of values) information. However, the very purpose of mobile crowdsensing is to allow collection of information that is *complex*. By complex, we mean information that is (i) *multimedial*, for example audio and video; and (ii) *qualitative*, which expresses an opinion regarding a particular event under consideration, for example, traffic/weather status over a particular area. Solving this issue is paramount to obtaining meaningful information from each sensing report and ultimately, assessing its QoI, but at the same, it is extremely challenging.

Multimedia input includes audio/video recording and pictures [10, 136], and are the most challenging to process in terms of QoI. Regardless, we need novel algorithms to (i) compute QoI on multimedia information; and (ii) compare QoI levels between different multimedia inputs if we want to apply existing truth discovery and trust mechanisms on this type of data. Although there exist some work in the context of sensor networks [41, 74], to the best of our knowledge the problem of QoI in multimedia-based mobile crowdsensing remains substantially unexplored. Among existing work, in [141, 144] we proposed trust frameworks for traffic monitoring applications. In particular, the use of MTPs could be fundamental to assess trustworthiness of participants in opinion-based systems. However, it is yet to be explored how to model and manage opinions. One possible direction could be to embed mechanisms similar to Jøsang's trust or fuzzy logic to model opinions.

### 7.0.2 Gamification for QoI Improvement.
The concept of gamification [38] has been proposed in many contexts as a viable and effective technique to increase QoI by persuading the participants with a natural (i.e., based on joy or entertainment) incentive type, instead of a remunerative one [91, 121, 169]. For example, the leaderboard-based system used by Waze [52] has proven to be effective in motivating the users to contribute reliable data. On the other hand, recent studies [66] cast doubts on gamification, as the authors argue that gamification effects are greatly dependent on the context in which gamification is being implemented, as well as on the users using it.

It is still uncertain whether these mechanisms may work effectively in mobile crowdsensing. In particular, we need to understand and validate the conditions by which gamification may result in a good strategy, and possibly, embed existing trust/truth discovery framework in the context of gamification. Also, gamification is currently based on heuristics, and we lack modeling languages for gamification [70] which may be able to capture the complex interactions in mobile crowdsensing.

### 7.0.3 Game-theoretical Trust Frameworks.
As one possible future research direction, we envision the development of game-theoretical trust frameworks based on *behavioral game theory* (BGT) [21].

Specifically, the concept of *bounded rationality* [159, 160] may be used to analyze less-than-rational behavior in participants. Below are some examples of why participants may not be perfectly rational:

- Participants may lack the ability to *understand* the game, for example, because of incomplete information at their disposal. The information may also be *asymmetrical*, which means, some users know more than others about the game being played.
- Participants may not be able to play the optimum strategy if the interaction between the user and the system is repeated. Also, the equilibrium played by the participants may be calculated by considering a different time-horizon, and therefore, mismatch the ones calculated by the system.
- In realistic contexts, the participating users might not know the details of the game been played. Therefore, they need to *learn* over time the strategy that maximizes the payoff. In such context, users might have different *learning strategies*. For example, some might use a reinforcement approach [11] (i.e., adapting the strategy as a function of current and past payoffs), while some might use anticipatory learning [22] to reason more thoughtfully on what the other players are doing and the game being played.

Therefore, it becomes necessary to embed these and other procedural aspects while modeling trust frameworks. We need to identify restrictions on the space of strategies and then optimize based on those restrictions. One way to incorporate procedural aspects, and therefore, uncertainty in the equilibria that will be played by users, is to study the *Quantal Response Equilibria* (QRE) of the game, a concept first proposed in [113]. QRE provide a statistical framework to analyze games that incorporate realistic limitations to rational choice modeling of games. In particular, QRE allow every strategy to be played with non-zero probability, and thus any strategy is possible (though not necessarily reasonable). By far, the most common specification for QRE is the *logit* equilibrium (LQRE) [113], where the player's strategies are chosen according to the probability distribution

$$\mathbb{P}\{a_i\} = \frac{exp\{\lambda \cdot \sum_{a_{-i} \in A_{-i}} \mathbb{P}\{a_{-i}\} \cdot u_i(a_i, a_{-i})\}}{\sum_{a'_i \in A_i} exp\{\lambda \cdot \sum_{a_{-i} \in A_{-i}} \mathbb{P}\{a_i\} \cdot u_i(a'_i, a_{-i})\}} \quad (23)$$

The $\lambda$ parameter is defined as the "rationality" parameter: the choice of action becomes purely random as $\lambda \to 0$, whereas the action with higher expected payoff is chosen as $\lambda \to \infty$. Analogously to Nash equilibria, QRE can be calculated for normal-form games [113] and extensive-form games [114]. Recent advances in the field have shown that the computation of QRE can be done by maximum likelihood estimation of the model's parameters, as explained in [31], or by techniques such as the homotopy method, developed in [149]. The value of $\lambda$ is usually chosen according to experimental data; however, such data does not exist yet to formulate QRE for games in the context of mobile crowdsensing.

Additional research is therefore needed to model the $\lambda$ parameter and therefore, obtain more realistic equilibria. Although it is true in general that irrationality might be dependent on the mobile crowdsensing application being considered, we believe that large-scale data collection campaigns with sufficient number of participants could help solve this issue.

## 8 CONCLUSIONS

Mobile crowdsensing has become one of the most promising paradigms for social sensing, as it allows to dramatically reduce infrastructure costs and obtain detailed information about the phenomenon being monitored. On the other hand, it brings into the picture novel and exciting multi-disciplinary research challenges, most of them are yet to be understood and solved. In particular, given the strong presence of human beings in the sensing loop, understanding in details the human dynamics and predictable behavior in mobile crowdsensing is fundamental to guarantee quality of information (QoI).

In this paper, we have surveyed the topic of QoI in mobile crowdsensing. After proposing a comprehensive framework that defines the concept of QoI in mobile crowdsensing, we have defined and analyzed in details the research challenges in enforcing and estimating the QoI, as well as discussing the already existing related work on the topic. By analyzing the limitations and weaknesses of existing work, we have provided a roadmap of possible directions of novel future research in the field.

The research challenges defined in this paper are just the tip of the iceberg. These and many other challenges, such as energy consumption, security and privacy, efficiency, and so on, are still to be tackled. Nevertheless, we expect that in the future, mobile crowdsensing applications will become deeply in our daily lives, and will help build a society in which sensing will be ultimately ubiquitous and protagonist in every aspect of human life.

## ACKNOWLEDGEMENT

We would like to thank the anonymous reviewers for their valuable comments, which helped us improve the quality of the paper. This material is based upon work supported by the National Science Foundation under Grant No. CNS-1545037, CNS-1545050, DGE-1433659 and IIS-1404673. Any opinions, findings, and conclusions or recommendations expressed in this material are those of the authors and do not necessarily reflect the views of the National Science Foundation.